# Long Rayleigh length confocal microscope: A fast evaluation tool for obtaining quantum properties of color centers

Yuta Masuyama,[1,*] Chikara Shinei,[2] Shuya Ishii,[1] Hiroshi Abe,[1] Takashi Taniguchi,[2] Tokuyuki Teraji,[2] & Takeshi Ohshima[1]

[1] National Institutes for Quantum Science and Technology, Takasaki, Gunma, 370 – 1292, Japan
[2] National Institute for Materials Science, Tsukuba, Ibaraki 305 – 0044, Japan
* masuyama.yuta@qst.go.jp

## Abstract

Color centers in wide band-gap semiconductors, which have superior quantum properties even at room temperature and atmospheric pressure, have been actively applied to quantum sensing devices. Characterization of the quantum properties of the color centers in the semiconductor materials and ensuring that these properties are uniform over a wide area are key issues for developing quantum sensing devices based on color center. In this article, we will describe the principle and performance of a newly developed confocal microscope system with a long Rayleigh length (LRCFM). This system can characterize a wider area faster than the confocal microscope systems commonly used for color center evaluation.

## Introduction

Color centers such as NV centers in diamonds are expected to be an important tool for quantum sensing because they are qubits that can be operated at room temperature and atmospheric pressure [1 - 4]. In addition to NV centers in diamond, various other color centers, such as silicon vacancies in SiC [5, 6], are beginning to be used depending on the sensing target application [7]. Magnetic sensing using quantum properties of NV centers has achieved the sensitivity of less than 1 pT/sqrt (Hz) for both AC magnetic field sensing [8] and DC magnetic field sensing [9] by applying a large sensor volume. Refining quantum sensor materials will make further sensitivity improvements possible [10]. In addition to magnetic field detection, the quantum sensor using color centers can measure temperature [11, 12], electric field [13], and pressure [14]. Furthermore, measurement techniques using multiple quantum sensors are also beginning to develop and improve sensor performance [15, 16].

One effective way to increase sensitivity is to improve the spatial uniformity of quantum properties within the sensor material. Confocal microscopy (CFM) and electron spin resonance (ESR) has been commonly used to evaluate quantum properties of color centers. Although CFM is a measurement technique with excellent spatial resolution, it is not suitable for materials with large sensor sizes because of small detection volume. Thus, the CFM is best suited for evaluating a small number of color centers, such as a single color center. However, measuring the spatial distribution of the entire sample by the CFM is impractical because of the long measurement time required due to the small detection volume. ESR is a method that evaluates the quantum properties of the entire sample as an average value and is not suitable for evaluating the spatial distribution of quantum properties. In order to obtain a guideline for improving sensitivity, a method that can evaluate the spatial distribution of quantum properties in the sensor materials within a millimeter-order spatial range is desired.

In this study, we describe a design protocol for an optical system that realizes a measurement system capable of evaluating a wide spatial distribution of quantum properties of color centers and their concentration while minimizing the effects of surface and background light. Here, we introduce a newly developed confocal microscope system that is featured by a long Rayleigh length (LRCFM) in an instrument configuration similar to a CFM system. By injecting a laser beam that is uniform in the depth direction and has a relatively larger spot size, the detection volume becomes significantly large. This effect makes it possible to evaluate the spatial distribution of quantum properties of color centers over the entire sample within a reasonable time. By optimizing the optical excitation technique, optical detection system, and microwave circuit, our proposed LRCFM method can evaluate color centers in a wide range of concentration ensembles, from about one ppb and above, up to high concentration color centers such as ten ppm, for example.

## Results and discussion

### Long Rayleigh length confocal microscope (LRCFM)

The LRCFM is characterized by increasing the excitation volume of the sample, thereby reducing the amount of unpolarized fluorescence that becomes noise. This increase in excitation volume can be achieved by exciting the entire thickness direction of the sample. Figure 1 shows a schematic diagram for comparison with existing color center evaluation methods. The spatial resolution, which correspond to the sample volume to be measured, of LRCFM is inferior than that of CFM while better than that of ESR. In the practical use, LRCFM is suitable for obtaining spatial distribution mapping of quantum properties across the entire mm-size sample in a realistic time. The proposed method polarizes spin states as efficiently as CFM. In addition, unlike CFM, this method is not affected so much by stray light of the material surface or background light because of the large excitation volume to be measured, resulting in high detection signal. Eventually, the measurement time becomes very short when using LRCFM. Details of features of LRCFM are described below.

|  | LRCFM (This study) | CFM | ESR (CW) | ESR (Pulse) |
|---|---|---|---|---|
| Schematics | Laser — Lens — Waist radius — Sample | Laser — Lens — Focal length — Sample | Sample, B magnetic field, Coil | Sample, Coil |
| Spatial resolution (plane) | ~ 20 μm | ~ 0.5 um | < 1mm | 1 mm < |
| Spatial resolution (depth) | ~ 500 μm | <5 um | < 1mm | 1 mm < |
| Sample volume for evaluation | ○ Small | ◎ Very small | × Large | × Large |
| Spatial mapping of quatum property | ○ Possible | △ Partially possible | × Impossible | × Impossible |
| Stray light | ○ Negligible | × Critical | ○ Negligible | ○ Negligible |
| Measurement time | ○ Short | × Long | × Long | × Long |

Fig. 1. Comparison of this study with confocal microscopy (CFM) and electron spin resonance (ESR).

**Rayleigh length and excitation volume**

First, the optical features of this study are described. Figure 2 (a) shows a schematic comparison of the beam diameter of LRCFM and CFM. CFM commonly uses lens system with high magnification objective lens. On the other hand, LRCFM with low magnification objective lens has a relatively larger laser beam diameter of about 20 μm and enables uniform color center excitation in the depth direction. The Rayleigh length is used as a measure of laser depth. The Rayleigh length $z_R$ is the length from the focal point to the point where the beam radius is $\sqrt{2}$ times the waist radius $w_0$ which is beam radius at the focal point of the laser. And the Rayleigh length is defined as

$$z_R = \frac{\pi w_0^2}{\lambda}, \tag{1}$$

where $\lambda$ is the wavelength of the laser [Fig. 2 (b)] [17]. Increasing the beam waist radius allows laser irradiation with a long Rayleigh length.

In this study, the calculations were made by approximating the laser excitation region as a cylindrical shape [the shaded area in Fig. 2 (c)]. The average power density of the laser light passing through this cylinder is $\frac{P_0}{\pi w_0^2}$, where $P_0$ is the total power of the laser. A 50x objective lens, which has an effective focal length of 3.6 mm, is commonly used in the case of CFM systems, as shown in Fig. 2 (c). When the twice Rayleigh length $2 z_R$ is shorter than the sample thickness, a cylinder of length (indicated by "Rayleigh length" in the Fig.2) $2 z_R$ and waist radius $w_0$ is taken as the excited volume in the calculation. On the other hand, when twice the Rayleigh length is longer than the thickness of the sample, the calculation was done using the thickness of the sample as the length of the cylinder. As the laser beam waist radius increases, the laser excitation volume rapidly increases. In the calculations below, the sample thickness of 500 μm and the incident 532 nm laser beam radius of 450 μm were applied.

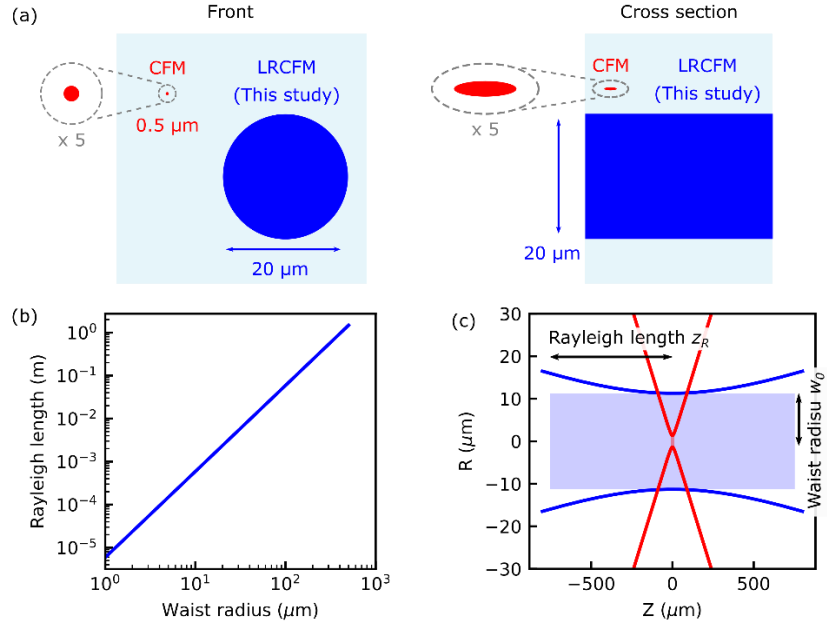

Fig. 2. (a) Schematic comparison of this study (LRCFM) and a confocal microscope (CFM) system. (b) Dependence of Rayleigh length on waist radius for laser wavelength 532 nm. (c) Excitation volume along the optical axis direction for focal lengths of 3.6 mm (red) and 30 mm (blue). The effective focal length of 3.6 mm corresponds to a 50x objective lens, and the focal length of 30 mm is used in this study. The blue and red colored squares with twice the Rayleigh length and twice the waist radius on each side represent the effective beam areas with focal lengths of 30 mm and 3.6 mm, respectively.

The amount of detected light was evaluated by the calculation steps shown in figure 3(a) to find the optimal laser beam condition for the LRCFM. Specifically, the excitation volume was calculated by changing the Rayleigh length of the excitation laser, and then the amount of fluorescence emitted by polarized color centers within this excitation volume and the amount of collected light were evaluated. Here, these calculations were performed using the energy levels of the negatively charged NV centers in the diamond [Figure 3(b)] (see supplemental material for the details). The values of the transition rates of the NV center required for this numerical calculation were taken from the reference by [18].

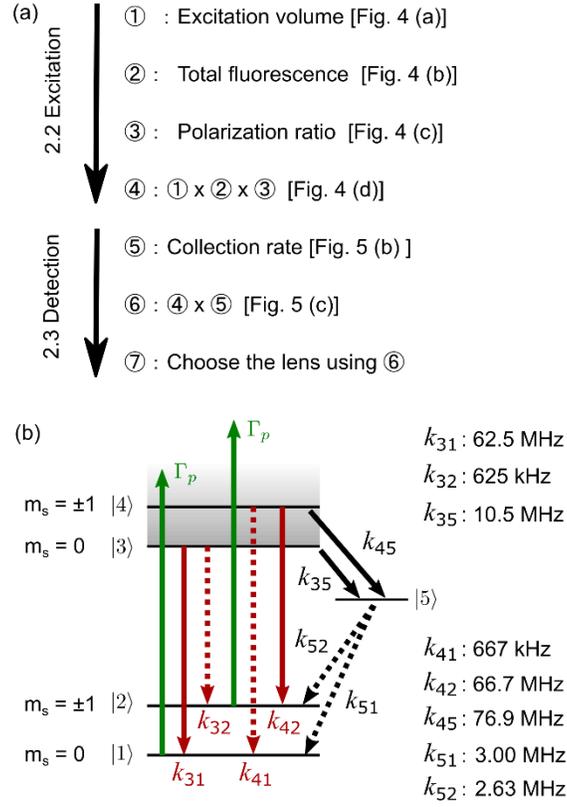

Fig. 3 (a) Calculation step to determine the focal length of the objective lens for LRCFM. (b) Energy level diagram for the NV center in diamond. The green arrow indicates optical excitation from the ground state of the NV center. The red arrow indicates fluorescence from the excited state of the NV center. $m_s$ is magnetic sublevel of the NV center. $k_{ij}$ indicates the transition rate from level i to level j.

**Excitation signal amount**

When the waist radius is very large, the Rayleigh length increases but the photon power density decreases greatly, resulting in a significant decrease in the polarization ratio of the color center. This phenomenon is confirmed from the numerically calculated amount of fluorescence emitted from the region excited by the laser beam. Figure 4 shows the calculated results for the NV center.

The amount of fluorescence emitted from the NV center is calculated using the steady-state population in states |3> and |4> as

$$I_{CW} = \frac{k_{31}+k_{32}}{k_{31}+k_{32}+k_{35}} \rho_{33}^{ss} + \frac{k_{41}+k_{42}}{k_{41}+k_{42}+k_{45}} \rho_{44}^{ss}, \quad (2)$$

where $\rho_{33}^{ss}$ and $\rho_{44}^{ss}$ are the elements of the steady state density matrix corresponding to the states |3> and |4> and $k_{ij}$ is the transition rate from level i to level j (see Appendix) [19]. Figure 4 (a) shows the numerical results of the Rayleigh length dependence of the excitation volume at a laser power of 10 mW. The solid blue line corresponds to a sample with a thickness of 500 μm, and the dashed red line corresponds to a sample with infinite thickness. As shown in Fig. 4 (b), the fluorescence per unit volume decreases due to the lowering of laser beam power density.

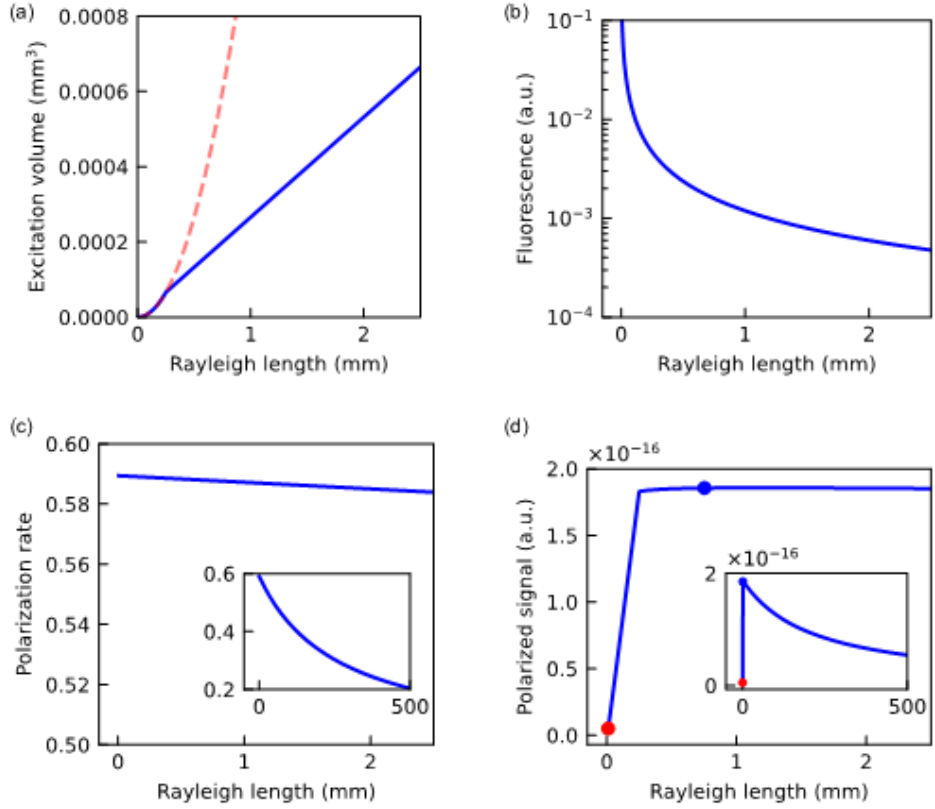

Fig.4 Numerical calculations for a 532nm laser with a beam diameter of 0.9 mm at a laser power of 10 mW. (a) Dependence of excitation volume on the Rayleigh length for sample thicknesses of 500 μm (solid blue line) and infinity (dashed red line). (b) Rayleigh length dependence of fluorescence intensity. (c) Rayleigh length dependence of polarization rate. (inset) A wider view of the Rayleigh length region in Fig. 4 (c). (d) Rayleigh length dependence of the product of excitation volume, fluorescence per unit volume, and polarization ratio. Blue and red circles correspond to the LRCFM and CFM cases, respectively. (inset) A wider view of the Rayleigh length region in Fig. 4 (d).

The polarization rate of a color center is an important indicator because a quantum state cannot be manipulated or read out unless it is polarized. The polarization rate of a quantum state is defined as

$$P = \frac{\rho_{11}^{ss} - \rho_{22}^{ss}}{\rho_{11}^{ss} + \rho_{22}^{ss}}, \tag{3}$$

where $\rho_{11}^{ss}$ and $\rho_{22}^{ss}$ are the elements of the steady state density matrix corresponding to the states |1> and |2>. In Fig. 4 (c), The polarization rate at a laser power of 10 mW is calculated. According to the Eq. (1), there is a positive correlation between the laser beam diameter and the Rayleigh length. The polarization ratio decreases with increasing beam diameter that can be written with the Rayleigh length because the laser beam power density becomes lower [Fig. 4 (c, inset)].

Furthermore, we study the total signal volume effective for quantum state readout. We calculated the dependence of the product of the detection volume ($= \pi \times w_0^2 \times$ sample thickness), the fluorescence per unit volume $I_{cw}$, and the polarization rate $P$ on the Rayleigh length at a laser power of 10 mW [Fig. 4 (d)]. The polarization ratio is good for the short Rayleigh length corresponding to the narrow waist radius, but the detection volume is so small that these product value becomes very small. The fluorescence intensity increases rapidly until twice the Rayleigh length reaches near the thickness of the sample, which corresponds to 500 μm. Then, as the twice the Rayleigh length increases further, the product's value decreases, mainly due to the decrease in laser power density [Fig. 4 (d, inset)]. Note that this calculation is for overall fluorescence emission. The focus calculation is given in the next subsection.

**Detection signal amount**

In the previous subsection, the total amount of emitted fluorescence was calculated which is represented by diagrams in Fig. 4. This subsection deals with the photon correction efficiency of the LRCFM system. Practically, we calculate the amount of detected fluorescence, taking into account the focusing performance of the objective lens [Fig. 5 (a)]. In this calculation, the beam radius of the incident laser is fixed at 450 μm, so the lens's focal length can be determined from the waist radius $w_0$. The numerical aperture (NA) was calculated from the focal length and the objective lens radius. The detection ratio was then calculated by comparing the ratio with NA = 1 [Fig. 5 (b)]. The amount of detected fluorescence for a laser power of 10 mW is obtained by multiplying the product of the detection volume (= $\pi \times w_0^2 \times$ sample thickness), the amount of fluorescence per unit volume $I_{cw}$, and the polarization ratio $P$. Fig. 5 (c) shows the detected fluorescence as a function of the Rayleigh length. This simulation shows that the maximum amount of the detected fluorescence occurs when twice the Rayleigh length matches the sample thickness. Our designed system can be used without concern for laser power because the Rayleigh length at maximum detected fluorescence is constant in the practical range of laser power (see supplemental material for the details).

Based on the results obtained so far, using the relation between the focal length of the objective lens and the waist radius $w_0 = 2\lambda F/\pi D$, where F is the lens's focal length and D is the diameter of the incident beam, the required focal length of the lens is obtained by

$$F = \frac{D}{2}\sqrt{\frac{z_R \pi}{\lambda}}. \tag{4}$$

This laser beam, focused by an objective lens with a focal length of F, excites the sample uniformly in-depth and produces a large amount of fluorescence. Because of the strong collected signal and larger beam spot size, typically about 20 μm, the LRCFM can measure the spatial distribution of properties throughout the entire mm-scale size sample in a realistic time.

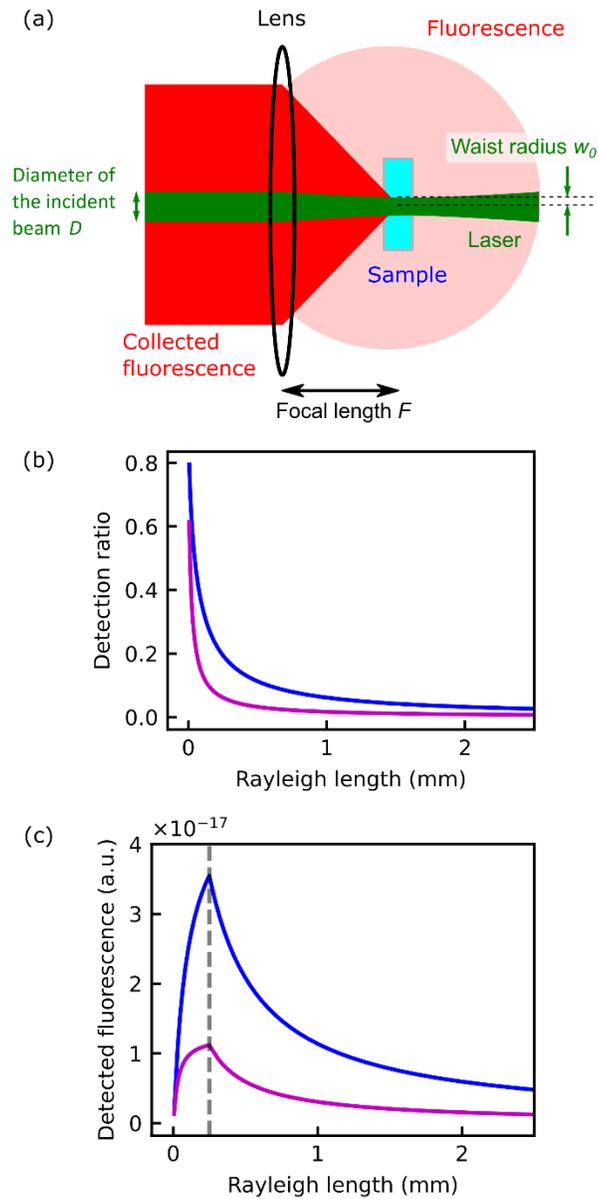

Fig. 5 (a) Schematic of a sample being excited by a laser and its fluorescence being collected. (b) Dependence of detection rate on excitation Rayleigh length for a 1-inch lens (blue) and a 0.5-inch lens (magenta). (c) Dependence of the product of the polarized fluorescence [Fig. 4 (d)] and the detection rate [Fig. 5 (b)] for a 1-inch lens (blue) and a 0.5-inch lens (magenta) on the excitation Rayleigh length at 10 mW laser power. The dashed line represents the location where the Rayleigh length is 0.25 mm, and twice this Rayleigh length equals the thickness of the sample.

### Choosing the objective lens and the optical fiber

In our experimental setup using LRCFM, the incident beam diameter of a 532 nm laser is 0.9 mm, and an objective lens with a focal length of 30 mm collects the most polarized signal when selected from a commercially available 1-inch achromatic lens. The 1-inch achromatic lens with a focal length of 30 mm was used in the following experiment.

In conventional CFM, the light collection efficiency increases due to the higher NA of the objective lens compared to LRCFM, at the same time, the spatial resolution improves. On the other hand, the amount of collected fluorescence decreases due to the effect of pinholes in the CFM, as indicated by red rectangle in Fig. 6 (a). The advantage of the LRCFM is that the measurement time is shortened by making identical size between the detection and the excitation regions, which corresponds to a detection proportion equal to one. The drawback is poor spatial resolution. A large pinhole, which corresponds to optical fiber core in the case of LRCFM, makes this large detection proportion possible, as indicated by red and green rectangles in Fig. 6 (a). The multimode fiber with a core diameter of 200 μm was used in the following experiment. By replacing the objective lens, optical fiber, and photodetector, it is possible to turn a conventional CFM into a LRCFM.

Compared to CFM with infinitely small pinhole diameters, the detected fluorescence of LRCFM is more than $10^4$ times higher [Fig. 6 (b)]. Thus, assuming that the noise levels of both systems are the same, CFM requires $10^8$ times the measurement time to achieve the same signal-to-noise ratio as LRCFM. In addition, the larger detection volume means that the surface occupies a tiny percentage of the total volume, making it less susceptible to the emission light from impurities and other substances on the surface. Due to its high photon focusing efficiency, LRCFM has a wide concentration range of detectable color centers.

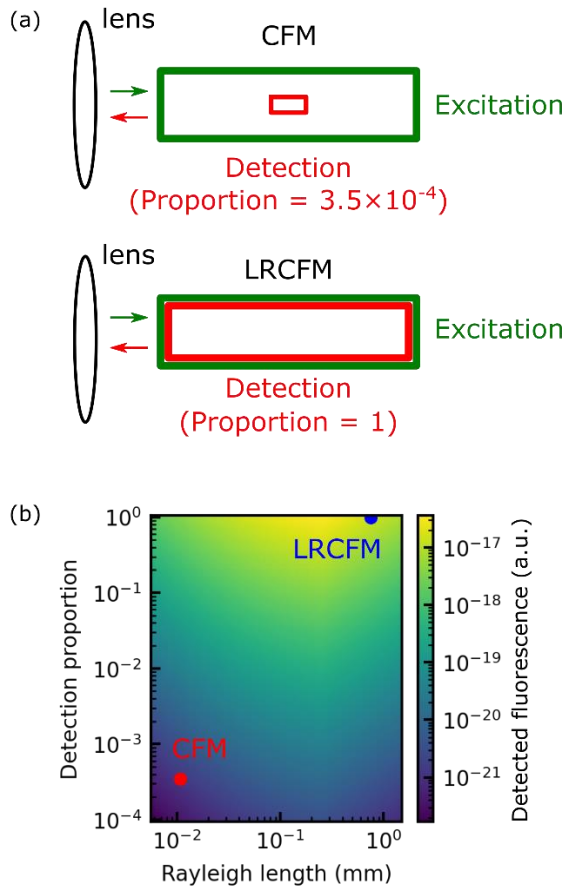

Fig. 6 (a) Schematic comparison of excitation and focusing volumes for confocal microscopy (CFM) and this study. (b) Product of detection rate [ Fig .4 (d)] and polarized fluorescence at 10

**Measurement of the spatial distribution of the quantum properties**

Figure 7 shows the spatial distribution of the quantum properties of NV centers in the diamond. One pixel size is 50 μm × 50 μm, and the spatial measurement area was up to 350 μm × 1050 μm. Here, the x and y coordinates are those shown in Fig. S1(b). Figure 7 (a) shows the results of Rabi oscillation measurements at the point (x, y) = (-200um, 100um) of Fig. 7(b), fitted by the function $a_1 \exp(-\tau/a_2) \cos(2\pi a_3 \tau + a_4) + a_5$, where τ is the microwave irradiation time, $a_i$ (i = 1 ~ 5) is a fitting parameter. Figure 7(b) shows the spatial distribution of the π pulse duration that inverts the quantum state. This data was used to correct for differences in π pulse duration due to different positions on the microwave circuit, thereby improving the accuracy of the $T_1$ and $T_2$ measurements. The π/2 pulse was set to half of the π pulse time. Figure 7 (c) shows the results of the energy relaxation time $T_1$ measurement at the point (x, y) = (-200um, 100um), fitted by the function $a_1 \exp(-\tau/a_2) + a_3$, where τ is the time duration between π pulse and readout, $a_i$ (i = 1 ~ 3) is a fitting parameter. In $T_1$ measurement, π pulses of microwaves were irradiated, and the quantum state was read out as a photoluminescence intensity by a 532 nm laser excitation after a time τ. Figure 7 (d) shows the spatial distribution of the energy relaxation time $T_1$. The average value over the entire measurement region of the sample was 11.0 ms with a standard deviation of 4.0 ms. Figure 7 (e) is the result of the phase relaxation time $T_2$ measurement at the point (x, y) = (-200um, 100um), fitted by the function $a_1 \exp(-(\tau/a_2)^{a_3})$, where τ is the time between the π pulse and the readout, $a_i$ (i = 1 ~ 3) is a fitting parameter. The microwave pulse sequence is π/2 pulse- π pulse - π/2, with π/2 pulse and π pulse separated by time τ. After an exposure of the microwave pulse sequence, the quantum states were read out with a 532 nm laser. Figure 7(f) shows the spatial distribution of the phase relaxation time $T_2$. The average value over the entire measurement region of the sample was 21.5 us and the standard deviation was 1.9 us. The value of $T_2$ is gradually decreasing from the left-side area to the right-side area. This trend of $T_2$ spatial distribution corresponds to the trend of the spatial distribution of fluorescence from the $NV^-$ center, as shown in Fig. S1(b). Fig. 7(f) and Fig. S1 (b) indicate that $T_2$ is decreasing in increasing density of nitrogen and $NV^-$ center as a decoherence source.

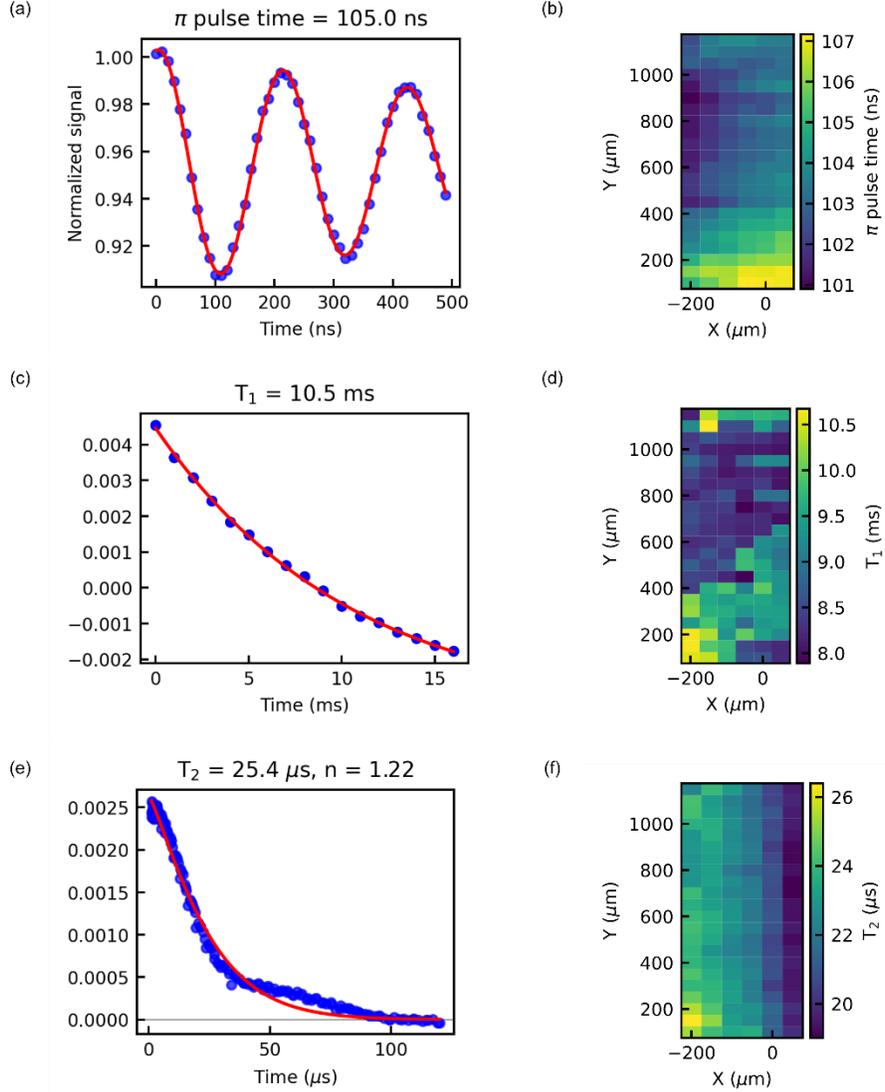

Fig. 7 (a) Rabi oscillation measurement at the point (x, y) = (-200um, 100um). (b) Spatial distribution of π pulse duration. (c) Energy relaxation time $T_1$ measurement at the point (x, y) = (-200um, 100um). (d) Spatial distribution of NV center energy relaxation time $T_1$. (e) Phase relaxation time $T_2$ measured at the point (x, y) = (-200um, 100um). (f) Spatial distribution of phase relaxation time $T_2$ at the NV center.

## Conclusion

In summary, we have developed a method to rapidly measure the sample information such as quantum properties of an entire mm-scale sample by constructing a confocal microscope with a long Rayleigh length. By considering the excitation volume, the polarization ratio of the quantum state, and the detection efficiency, we found that matching twice the Rayleigh length to the thickness of the sample produces the highest amount of detected signal. The detected signal is about $10^4$ times larger than the conventional ideal confocal microscopy. LRCFM uses an identical microwave setup with conventional confocal microscopy to evaluate color centers.

Thus, it is possible to use these microwave pulse sequences for LRCFM, such as evaluating NV center density using instantaneous diffusion [22, 23]. Density evaluation of spin defects in the quantum material, usually performed on the entire sample by ESR [24], can be accomplished using LRCFM, with additional information on spatial distribution. In addition, this method can be used not only for the evaluation of existing color centers, but also for the discovery of new color centers, as it can take advantage of the high signal-to-noise ratio to increase the speed for measuring photoluminescence spectra of unknown color centers. [25, 26].

## Method

A diamond sample containing NV centers was placed on the microwave resonator [20]. The resonator is fixed to a motorized stage with a movable range of 13 mm. A permanent magnet to fix the quantization axis was placed behind the resonator. The strength of the magnetic field was approximately 2.5 mT. The fluorescence from the NV centers was collected through the objective lens (AC254-030-AB-ML; Thorlabs) and detected using an avalanche photodiode (APD410A/M; Thorlabs). An oscilloscope (MDO34; Tektronix) recorded the signal from the avalanche photodiode and analyzed it to distinguish the spin state of the NV centers. Using an arbitrary waveform generator (M3202A; Keysight) to generate the laser and microwave pulse sequence, we controlled an acousto-optic modulator (#35 250-0.2-0.53-XQ; Gooch & Housego) and a transistor-transistor logic (TTL) switch. After the microwave pulse sequence, a 180 µs measurement laser pulse of 532 nm laser (gem 532; Laser Quantum) for measurement with 12 mW was irradiated to the NV centers in the diamond. The diamond {111} single crystal grown by high-temperature/high-pressure (HPHT) synthetic method was used in this study [21]. The HPHT sample was irradiated with a 2.0 MeV electron beam with a total fluence of $5.0 \times 10^{17}$ cm$^{-2}$ for creating vacancies in the crystal and then annealed at 1000 °C for 2 hours in vacuum to create NV centers (see supplemental material for the details).

## Data availability

The data that support the findings of this study are available from the corresponding author upon reasonable request.

## Acknowledgements

This work was supported by MEXT Q-LEAP (JPMXS0118067395 and JPMXS0118068379). YM acknowledges the support of JSPS KAKENHI (20K14392). T.T. acknowledges the support of JST Moonshot R&D (JPMJMS2062), MIC R&D for construction of a global quantum cryptography network (JPMI00316) and JSPS KAKENHI (20H02187, 20H05661).

## Author information

### Authors and Affiliations

**National Institutes for Quantum Science and Technology, Takasaki, Gunma,**

**370 – 1292, Japan**

Y. Masuyama, S. Ishii, H. Abe & T. Ohshima

**National Institute for Materials Science, Tsukuba, Ibaraki 305 – 0044, Japan**

C. Shinei, T. Taniguchi & T. Teraji

### Contributions

The measurement system was designed, constructed, and tested by Y.M. Y.M. and C.S. performed the measurements and the data analysis. T.T. synthesized the diamond. The diamond was electron-irradiated by S. I. and H.A. The overall supervision was performed by T.T. and T.O.

### Corresponding author

Correspondence to Yuta Masuyama

## Ethics declarations

### Competing Interests

The authors declare no conflicts of interest.